# Dielectric Relaxation and Electrical Conductivity in $Bi_5NbO_{10}$ Oxygen Ion Conductors Prepared by a Modified Sol-Gel Process

Jungang Hou,[a,b] Rahul Vaish,[a,c] Yuanfang Qu,[b] Dalibor Krsmanovic,[a] K. B. R. Varma,[c] R. V. Kumar[a*]

[a]Department of Materials Science and Metallurgy, University of Cambridge, Cambridge CB2 3QZ, United Kingdom.
[b]Key Laboratory for Advanced Ceramic and Machining Technology of Ministry of Education, Tianjin University, Tianjin, 300072, China.
[c]Materials Research Centre, Indian Institute of Science, Bangalore-560 012, India.

**Abstract**

Crystalline $Bi_5NbO_{10}$ nanoparticles have been achieved through a modified sol-gel process using a mixture of ethylenediamine and ethanolamine as a solvent. The $Bi_5NbO_{10}$ nanoparticles were characterized by X-ray diffraction (XRD), differential scanning calorimetry/thermogravimetry (DSC/TG), fourier transform infrared spectroscopy (FT-IR), transmission electron microscopy (TEM) and Raman spectroscopy. The results showed that well-dispersed 5-60 nm $Bi_5NbO_{10}$ nanoparticles were prepared through heat-treating the precursor at 650 °C and the high density pellets were obtained at temperatures lower than those commonly employed. The frequency and temperature dependence of the dielectric constant and the electrical conductivity of the $Bi_5NbO_{10}$ solid solutions were investigated in the 0.1 Hz - 1 MHz frequency range. Two distinct relaxation mechanisms were observed in the plots of dielectric loss and the imaginary part of impedance (Z") versus frequency in the temperature range of 200-350 °C. The dielectric constant and the loss in the low frequency regime were electrode dependent. The ionic conductivity of $Bi_5NbO_{10}$ solid solutions at 700 °C is 2.86 $(\Omega \cdot m)^{-1}$ which is in same order of magnitude for $Y_2O_3$-stabilized $ZrO_2$ ceramics at same temperature. These results suggest that $Bi_5NbO_{10}$ is a promising material for an oxygen-ion conductor.

Corresponding author.
Phone: +44 1223-334327; Fax: +44 1223 334567
Email address: rvk10@cam.ac.uk

# 1. Introduction

Oxide ion conductors that exhibit the basic cubic fluorite structure are amongst the best known solid electrolytes and include compounds such as yttria stabilised zirconia and $\delta$-$Bi_2O_3$. The high-temperature formation of bismuth oxide $\delta$-$Bi_2O_3$, which has an oxygen-deficient fluorite-type structure, was recognized as one of the best solid-state oxide ion conductors owing to the high concentration of intrinsic oxygen vacancies. Solid electrolytes based on stabilized $\delta$-$Bi_2O_3$ with high oxygen ionic conductivity are of particular interest for the potential applications in electrochemical devices such as high-purity oxygen generators, electrochemical sensors and anode materials in solid oxide fuel cells [1-4]. Among the large number of choices of cations that can be used to stabilize cubic $\delta$-$Bi_2O_3$, Nb ion is one of the most frequently used owing to its highly efficient stabilization of the cubic phase down to room temperature. The minimum concentration of niobium oxide necessary to stabilize $\delta$-$Bi_2O_3$ is 10 mol% only [5]. With the continuous demand for miniaturization of electronic devices, the integration of $\delta$-$Bi_2O_3$-based devices into the new generation of electronic components represents a high priority.

Different kinds of techniques have been developed to prepare bismuth-based materials, such as magnetron sputtering, chemical solution deposition and pulsed laser deposition [6-8]. Conventional solid-state reaction method has been widely used to fabricate oxide solid solutions, in which high sintering temperature and long sintering time were required to achieve high density and homogeneous compositions. In comparison with conventional solid-state method, wet-chemical routes have several advantages, such as low processing temperature, the ability to provide better control of composition, homogeneity, purity, and so forth. The sol–gel method has been employed to fabricate various nanoparticles [9-10]. In the present work, well dispersed $Bi_5NbO_{10}$ nanoparticles have been achieved through a general aqueous sol-gel process with the assistance of a mixture of ethylenediamine and ethanolamine as a solvent. This general aqueous sol-gel process is based on the formation of organic-inorganic precursors obtained from metal-chelated by ethylene diamine tetra-acetic acid (EDTA) and citric acid. The EDTA/citric acid complexing process combined with the aid of the solvothermal route can lead to homogeneous mixing, better stoichiometric control, elimination of intermediate grinding steps and relatively low temperature annealing.

Though, the XRD studies of $Bi_5NbO_{10}$ have been studied in detail [11], the literature on the electrical properties is scarce. It deserves much attention from its dielectric properties point of view as these properties have direct influence on its electrical transport characteristics. The dielectric behaviour and high temperature electrical condcutivity were investigated. The effect of electrodes on the dielectric properties is discussed. These results will give some light on the mechanism of the electrical conduction.

## 2. Experimental

### 2.1 Synthesis

The following materials and reagents were used: bismuth nitrate ($Bi(NO_3)_3 \cdot 5H_2O$, Aldrich), niobium oxide ($Nb_2O_5$, Aldrich), hydrofluoric acid (HF, Aldrich), citric acid anhydrous ($C_6H_8O_7 \cdot H_2O$, Aldrich), ethylene diamine tetra-acetic acid, ethylenediamine, ethanolamine, ammonia solution (30%, Aldrich).

To begin with, niobium oxide ($Nb_2O_5$) was dissolved in HF to obtain a clear solution of the niobium-fluoride complex. The hydrous niobium oxide ($Nb_2O_5 \cdot nH_2O$) was then precipitated out from the clear solution of the niobiumfluoride complex by the addition of aqueous solutions of dilute (25%) ammonia. The precipitate of $Nb_2O_5 \cdot nH_2O$ was filtered and washed with 5% ammonia solution to make it fluoride free. The required amount of hydrous niobium oxide was then taken and slowly dissolved in aqueous solution of citric acid to obtain the niobium-citrate stock solution (solution A).

The Bi-EDTA complex solution (solution B) was prepared by dissolving stochiometrically calculated amounts of bismuth nitrate and EDTA in deionized water by stirring at room temperature, forming a colloid of Bi-species hydrolyzed with ammonia solution. The molar ratio of EDTA : CA : the total metal ions was 1 : 1 : 1. Then mixing the solution A and B, a mixture of ethylenediamine and ethanolamine with a molar ratio of 1 : 1 was added to the mixture solution. A small amount of aqueous ammonium hydroxide (28wt%) was added drop-wise to increase the pH value which was adjusted in the range of 6-9 by the addition of supplementary $NH_3 \cdot H_2O$, and the solution was then diluted to 0.1M. The resultant solution was gradually heated up to 60 °C and maintained over night for 12 h with mild magnetic agitation to aid gelation. The gel was prepared by evaporating water from the $Bi_5NbO_{10}$ sol at 100 °C in an electrical oven. As the viscosity of solution increased with dehydration, a dried gel started to form. After heat treatment of the gel at different annealing temperatures for 2 h, crystalline $Bi_5NbO_{10}$ nanoparticles were obtained. The annealed nanoparticles were dispersed in ethanol by ultrasonic atomization, dried, and mixed with 3wt% PVA, and then pressed into disks in a steel die at a pressure of 50 MPa, after which the pellets were finally pressed again using iso-static pressing at a pressure of 200 MPa. The disk-shaped samples were heated at 500 °C to remove the binder and then sintered at 800-900 °C to produce $Bi_5NbO_{10}$ solid solutions.

## 2.2 Characterization

After degassing/evaporation, solid products were stored in crucibles with a lid containing a small hole. Phase identification of the as-prepared samples was preformed using XRD through a Scintag diffractometer, operating in the Bragg configuration, using Cu $K_\alpha$ radiation. Samples were obtained by grinding thoroughly using a mortar and pestle. Diffraction patterns were collected from 10 to 90° at a scanning rate of 2° per minute with a step size of 0.02°. Parameters used for slit widths and accelerating voltages were identical for all samples. The voltage and current settings of the diffractometer were 40 kV and 40 mA, respectively. The particle size and size distribution were confirmed by transmission electron microscopy. The thermal decomposition behaviour of the samples was investigated by DSC–TG with a heating rate of 10 °C·min$^{-1}$ in a continuous air flow (100 mL·min$^{-1}$). FT-IR spectra were obtained by a FT-IR spectrophotometer from 350 to 4000 cm$^{-1}$ by the KBr pellet method. Raman spectra was measured by Raman spectrometer with a JY H-1 double monochromator and a photon counter under 633 nm laser excitation. Impedance studies were performed on fired-on-silver (750 °C for 0.5 h)/sputtering silver/gold electroded sintered samples using Impedance Analyzer (Solartron, SI 1260) in the 0.1 Hz-10 MHz frequency range at various temperatures. Besides the samples studied on the effect of different electrodes, the fired-on-silver for nomal samples was used as electrodes.

## 3. Results and discussion

The XRD patterns of the particles calcined at different temperatures through the EDTA/citric acid complexing process combined with the aid of the solvothermal route, are shown in Fig. 1. For the as-synthesized precursors, it can be observed that the materials calcined at 450 °C are amorphous. After being heated at 650 °C for 2 h, single $Bi_5NbO_{10}$ phase which is in agreement with the XRD patterns reported previously, begins to form and all the characteristic peaks of $Bi_5NbO_{10}$ phase appear without any second phase, whereas complete phase formation occurs only at 800 °C in the solid-state method [11,12].

**Figure 1.**

TEM was used to estimate the shape and particle size of $Bi_5NbO_{10}$ powders. Fig. 2a and 2b show typical TEM images of the as-synthesized nanoparticles. Majority of the primary crystallites have a size distribution within the range 5-100 nm as shown from laser granulometry in Fig. 2c and 2d. As shown in Fig. 2a and 2c, the sphere nanoparticles were formed at 650 °C, and their particle size was smaller than 60 nm. A size distribution range of 5 to 60 nm is obtained by laser granulometry experiment. After 850 °C, the particle size of the sphere nanopowders increased up to a range of 10-100 nm, demonstrating an average particle size of 60 nm with the standard deviation of 5 nm in Figure 2d. Obtaining such relatively fine nanoparticles by the modified sol-gel process has marked advantages over other chemical synthesis techniques.

**Figure 2.**

To clarify the chemical reaction during the thermal decomposition of the precursors, typical

TG-DSC curves for the crystallization of the precursors obtained at a heating rate of 10 °C·min$^{-1}$ in a continuous air flow are shown in Fig. 3. The sample weight decreases with increasing temperature continuously from room temperature to 350 °C, associated with the weight loss of 43 wt% as shown in TG curve. No significant weight loss can be observed from 350 to 1000 °C. There were two exothermic peaks at 208 and 264 °C, which can also be ascribed to the pyrolysis of the Bi-Nb organic chelating, combustion of nitrate and the organic phases (EDTA, CA, ethylenediamine and ethanolamine) to give an amorphous inorganic phase. The exothermic peak at about 600 °C is believed to be due to phase transition from amorphous oxides to $Bi_5NbO_{10}$ and impurities according to the XRD pattern (see Fig. 4). This transition process takes place simultaneously with the combustion of residual organic products and/or carbon [13], which is presumably caused by the fast heating rate. After this reaction, the remaining species consist of oxides of the Bi and Nb metal ions.

**Figure 3./ Figure 4.**

Fig. 5 shows the FT-IR spectra of the powders calcined at various temperatures in the wave number range between 400 and 4000 cm$^{-1}$. The spectrum of the precursor clearly shows a broad absorption around 3450 cm$^{-1}$, which is a charateristic stretching vibration of water and hydroxyl groups (O-H). Peaks localized at 1636 and 1395 cm$^{-1}$ are assigned to asymmetrical and symmetrical stretching vibration of carboxylate (O–C=O), which become weaker with increase in the annealing temperature. In addition, the spectrum (a) has the strongest peaks at 1636 and 1395 cm$^{-1}$ which is associated with the organics from the gel.

**Figure 5.**

The broad peak at 400–1000 cm$^{-1}$ for the samples have evolved into several peaks at 493, 586 and 870 cm$^{-1}$, after annealing due to the M–O vibrations. The spectra show the position of the cations in the crystal structure with oxygen ions and their vibration modes. These three bands can be ascribed to the stretching vibration of Nb–O band. There is no any peculiar peak is observed (as shown in inset of Fig. 5) for metal–metal and metal–oxygen networks within the amorphous phase which is also consistent with the XRD patterns. This clearly points to the existence of different arrangements of metal–oxygen–metal networks in these samples, depending upon the annealing temperature.

Fig. 6 displays the Raman spectra of powders annealed at temperatures ranging from 450 to 850 °C, investigating the short range structure information. The broadness of the features, probably due to the nanometric particles, prevents an accurate estimation of the precise band frequencies. It can be observed that the spectra underwent some modifications as the temperature rose. After analyzing the previous work on Raman spectra of pyrochlores [14-16], the typical bands of $Bi_5NbO_{10}$ powders were found to be located at 157, 268, 306, 505, 611 and 752 $cm^{-1}$ from 450 to 850 °C. The band at 157 $cm^{-1}$ was attributed to the Bi–O stretching modes, while the band at 752 $cm^{-1}$ was attributed to the Nb–O stretching mode and this intensity increased with increase of annealing temperatures which indicates the crystallization of $Bi_5NbO_{10}$ powders. It's possible that the bands at 505 and 611 $cm^{-1}$ were to correspond to the O–Nb–O bond bend. Besides in the spectra (a), no peaks have appeared in the sample annealed at 450 °C, which is interpreted as a signal of the nanoparticles being amorphous, in agreement with XRD measurement.

**Figure 6.**

The variation of the dielectric constant ($\varepsilon_r^{'}$) with the frequency (0.1 Hz-1 MHz) of measurement for $Bi_5NbO_{10}$ solid solutions at different temperatures is shown in Fig.7 (a). The dielectric constant decreases with increase in frequency. The decrease is significant especially at low frequencies which may be associated with the electrode polarization. The low frequency dispersion of $\varepsilon_r^{'}$ gradually increases with increase in temperature due to an increase in the electrode polarization as well as thermal activation of defect dipoles in the samples.

**Figure 7.**

When the temperature rises, the dispersion region shifts towards higher frequencies due to thermal activation of charge carriers. As the frequency increases, $\varepsilon_r^{'}$ decreases due to high periodic reversal of the field, which reduces the contribution of the charge carriers towards the dielectric constant. The variation of the dielectric loss (tanδ) with the frequency at various temperatures is shown in Fig. 7(b). Relaxation peaks were encountered in 200-350 °C temperature range. To investigate the origin of observed relaxation peak, different electrode materials (sputtered silver, sputtered gold and fired silver) were used. Figs. 8 (a & b) shows the dielectric constant and the loss behaviour at 250 °C for various electrode materials. Significant difference in the dielectric constants for different electrode materials was observed at low frequencies [Fig. 8(a)]. The discrepancy in dielectric relaxation is due to the fact that the sputtered nanosized metal clusters have a better contact on the surface of the sample than that of the fired silver, which has very large particles. Also, among the various other factors, the work function of the electrode material, thickness of the Schottky barrier, etc. play a crucial role. However, all the plots merge in the high frequency regime (above 3 kHz).

This electrode independent behavior at high frequencies (3 kHz-1 MHz) is attributed to the intrinsic dielectric response of the samples. Interestingly, clear relaxation peaks were observed in the frequency dependent dielectric loss plots [Fig. 8(b)]. The frequency associated with the dielectric relaxation was found to vary with the electrode materials used suggesting that the above relaxation is ascribed to electrode polarization.

**Figure 8./ Figure 9.**

In order to have further insight into the dielectric behavior of $Bi_5NbO_{10}$ solid solutions, the imaginary parts of impedance (Z") were plotted against frequencies at various temperatures (Fig. 9). The clear relaxation peaks were observed and shifted to higher frequencies with increase in temperature.

The activation energy involved in the relaxation process could be obtained from the temperature dependent frequency associated with the peak of Z" as:

$$f_m = f_o \exp(-E_R/kT) \qquad (1)$$

where $E_R$ is the activation energy associated with the relaxation process, $f_o$ is the pre-exponential factor, *k* is the Boltzmann constant and *T* is the absolute temperature. An inset of Fig. 9 shows a plot between ln($f_m$) and 1000/*T* along with the theoretical fit (solid line) to the above equation (Eq. 1). The value that is obtained for $E_R$ is 0.85 ± 0.03 eV, which is ascribed to the motion of oxygen ions and is consistent with the one reported in the literature for oxygen ion diffusion in oxide solid solutions [17]. The relaxation time (1/2π$f_o$) at infinite temperature for the electrical relaxation is 4 × $10^{-14}$ s, suggesting a mechanism of short range diffusion of oxygen ions via vacancies. Fig. 10 shows the normalized plots of impedance Z" versus frequency (for fired-on-silver electroded samples) wherein the frequency is scaled by the peak frequency. A perfect overlapping of all the curves on a single master curve is evidently noticed. The perfect time-temperature superimposition shows that the conduction mechanism remains unchanged with temperature. Fig. 11 depicts normalized plots for imaginary parts of impedance for different electrodes at 250 $^oC$. It is interesting to note that relaxation peaks were completely overlapped, indicating that relaxation mechanism is invariant with the electrodes.

**Figure 10./Figure 11.**

In order to further elucidate the transport mechanism in $Bi_5NbO_{10}$ solid solutions, ac conductivity at different frequencies and temperatures, was determined by using the dielectric data as the following formula

$$\sigma(\omega) = \omega \cdot \varepsilon_o \cdot \varepsilon" \tag{2}$$

where *σ(ω)* is the ac conductivity at a frequency *ω* (*ω* = 2π*f*) and $\varepsilon_o$ is the permittivity of free space. The frequency dependence of the ac conductivity at different temperatures is shown in Fig. 12 (for fired-on-silver electroded samples). At low frequencies, the conductivity shows a flat response which corresponds to the dc part of the conductivity. At higher frequencies, the conductivity shows a dispersion. It is clear from the figure that the flat region increases with the increase in temperature. In Fig. 12, at all the temperatures, the conductivity is independent of frequency at low frequency regime. The conductivity increases with increasing temperature due to thermal activation of conducting species in the samples. Electrical conduction in $Bi_5NbO_{10}$ solid solutions is expected to result mainly from the defects present in the lattice. The variation of dc conductivity ($\sigma_{dc}$) with temperature can be described by Arrhenius equation

$$\sigma_{dc} = \sigma_o \exp(-E_{dc}/kT) \qquad (3)$$

where $\sigma_o$ is pre-exponential factor and $E_{dc}$ is activation energy associated with dc conductivity. Fig. 13 shows dc conductivity as a function of inverse of absolute temperature. From the slope of the linear fit, we can estimate activation energy associated with dc conduction. The activation energy calculated from the slope of the fitted line is found to be 0.94 ± 0.02 eV. The value of activation energy is in close agreement with that of $Y_2O_3$-stabilized $ZrO_2$ ceramics [18]. The activation energy for dc conduction is sum of two contributions of the detrapping energy (0.85 eV) and the energy attributed to the mobility of the free ions (0.09 eV). It suggests that the ionic conduction in the $Bi_5NbO_{10}$ solid solutions takes place via hopping mechanism.

**Figure 12./Figure 13.**

## 4. Conclusion

In this study, well dispersed $Bi_5NbO_{10}$ nanoparticles with a particle size range of 5-100 nm have been achieved through a modified sol-gel process using a mixture of ethylenediamine and ethanolamine as a solvent. Results of XRD and DSC/TG analysis indicate that pure $Bi_5NbO_{10}$ phase was obtained above 650 °C. The variation in peak intensities of FT-IR and Raman spectra are useful for investigating the effect of annealing on the crystallization process. Thus, a significant feature of this approach is the large-scale quantities of well-dispersed $Bi_5NbO_{10}$ nanoparticles, which can also be extended to related materials, greater control over crystal morphology and tunability of dielectric and transport properties may be realized.

In particular, the ac conductivity and dielectric properties of $Bi_5NbO_{10}$ solid solutions were investigated as a function of frequency. The high value of dielectric constant at low frequencies was attributed to the electrode-related polarizations. The activation energy associated with the dielectric relaxation determined from the impedance spectra was found to be 0.85 ± 0.03 eV which is ascribed to the relaxation of oxygen ions. The activation energy for dc conductivity was 0.94 ± 0.02 eV. This suggests that the movements of oxygen ions are responsible for both ionic conduction as well as the relaxation process. The ionic conductivity of $Bi_5NbO_{10}$ solid solutions at 700 $^o$C is 2.86 $(\Omega\cdot m)^{-1}$ which is in same order of magnitude for $Y_2O_3$-stabilized $ZrO_2$ ceramics at same temperature. The electrical conductivity data indicated that the $Bi_5NbO_{10}$ solid solutions is good oxygen ion conductor and has potential applications in solid oxide fuel cells.

## Acknowledgements

This work was supported in part by EPSRC and PCME Ltd. The authors gratefully acknowledge the CSC (China Scholarship Council), Government of People's Republic of China, for providing financial assistance.

**Figure Captions**

**Fig. 1** (a) XRD patterns of $Bi_5NbO_{10}$ powders annealed at different temperatures for 2 h: (a) 450 °C (b) 650 °C, (c) 750 °C and (d) 850 °C.

**Fig. 2** Typical TEM images of the as-synthesized $Bi_5NbO_{10}$ nanoparticles annealed at 650 °C (a) and 850 °C (b), corresponding the laser granulometry curve from the samples treated at 650 °C (c) and 850 °C (d). Cumulative volume percentage as a function of particle size.

**Fig. 3** DSC/TG analysis curves of the as-synthesized $Bi_5NbO_{10}$ precursors.

**Fig. 4** XRD pattern of $Bi_5NbO_{10}$ powders annealed at 600 °C for 2 h.

**Fig. 5** FT–IR spectra of the as–synthesized $Bi_5NbO_{10}$ powders at different temperature for 2 h: (a) 450 °C, (b) 650 °C, (c) 750 °C and (d) 850 °C.

**Fig. 6** Raman spectra of $Bi_5NbO_{10}$ powders annealed at different temperatures: (a) 450 °C, (b) 650 °C, (c) 750 °C and (d) 850 °C.

**Fig.7** Frequency dependent behaviour of (a) dielectric constant and (b) dielectric loss plots at various temperatures.

**Fig. 8** Frequency dependent behaviour of (a) dielectric constant and (b) dielectric loss using various electrode materials at 250 °C.

**Fig. 9** Imaginary parts of the impedance as a function of frequency at various temperatures and the inset shows Arrhenius plot for electrical relaxation

**Fig. 10** $Z"/Z"_{max}$ versus $f/f_{max}$ plots at various temperatures for $Bi_5NbO_{10}$ solid solutions.

**Fig. 11** $Z"/Z"_{max}$ versus $f/f_{max}$ plots for various electrodes for $Bi_5NbO_{10}$ solid solutions.

**Fig. 12** Variation of ac conductivity as a function of frequency at different temperatures for $Bi_5NbO_{10}$ solid solutions.

**Fig. 13** Arrhenius plot for dc conductivity for $Bi_5NbO_{10}$ solid solutions.

**Figure 1.**

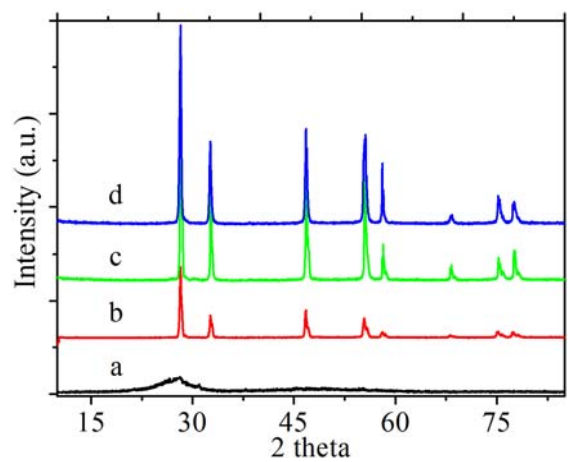

**Figure 2**

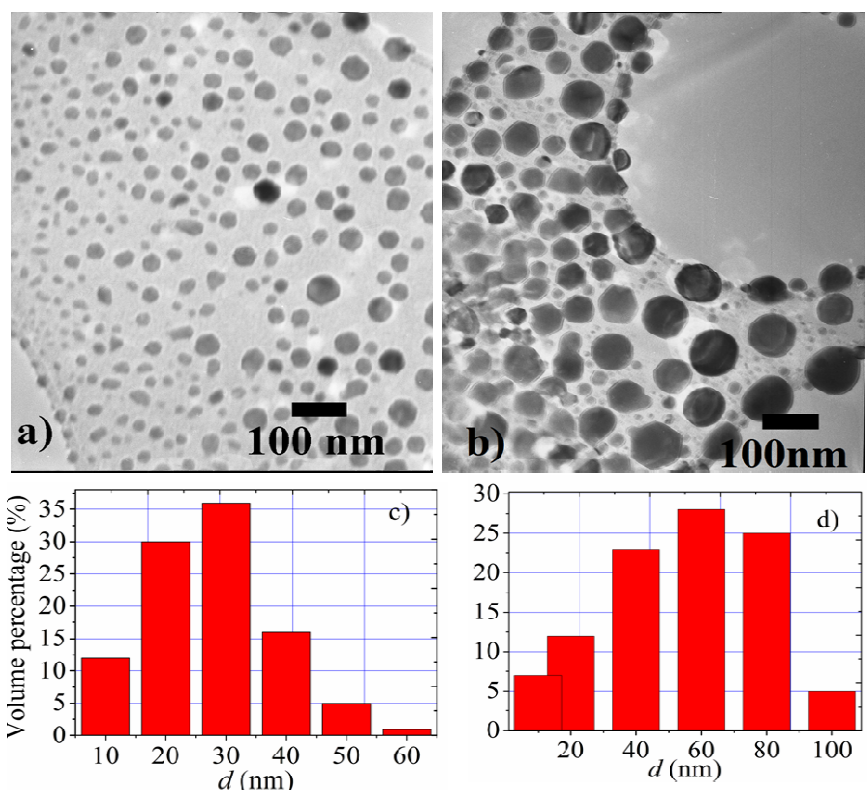

**Figure 3.**

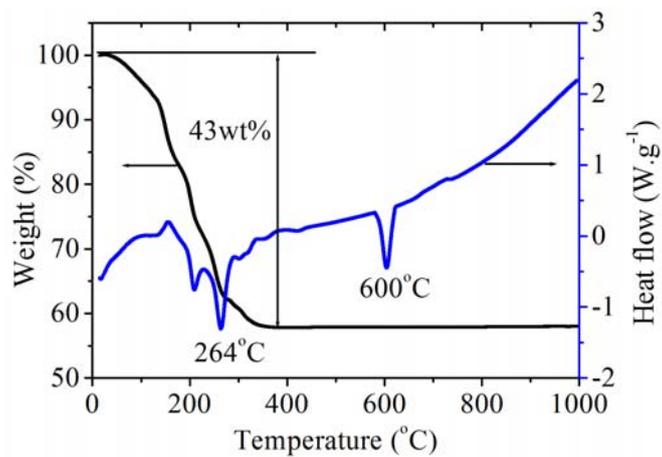

**Figure 4.**

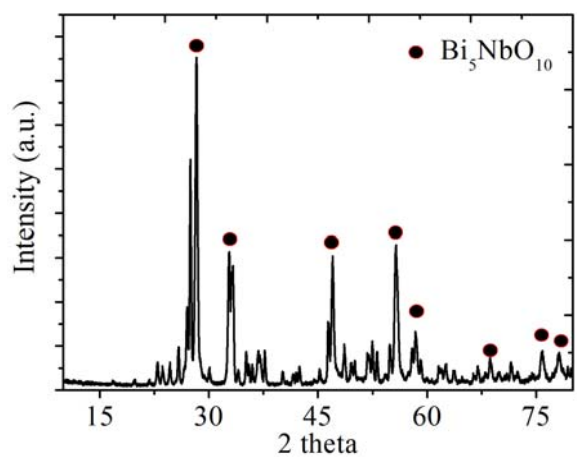

**Figure 5.**

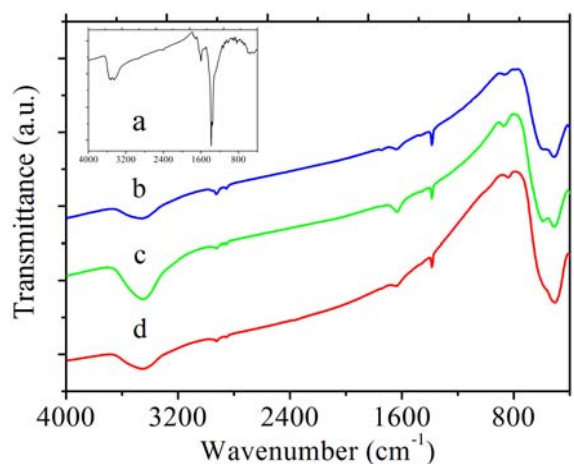

**Figure 6.**

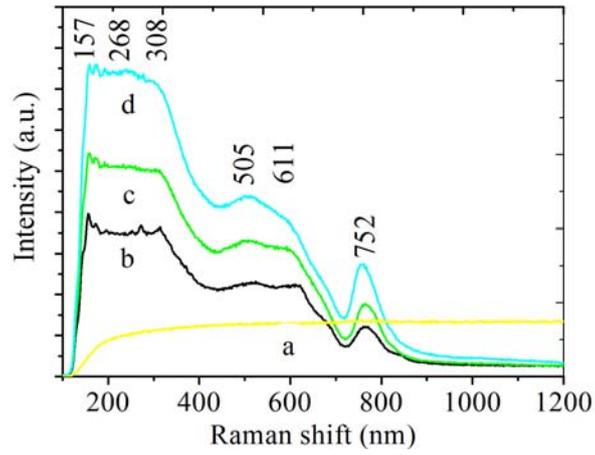

**Figure 7.**

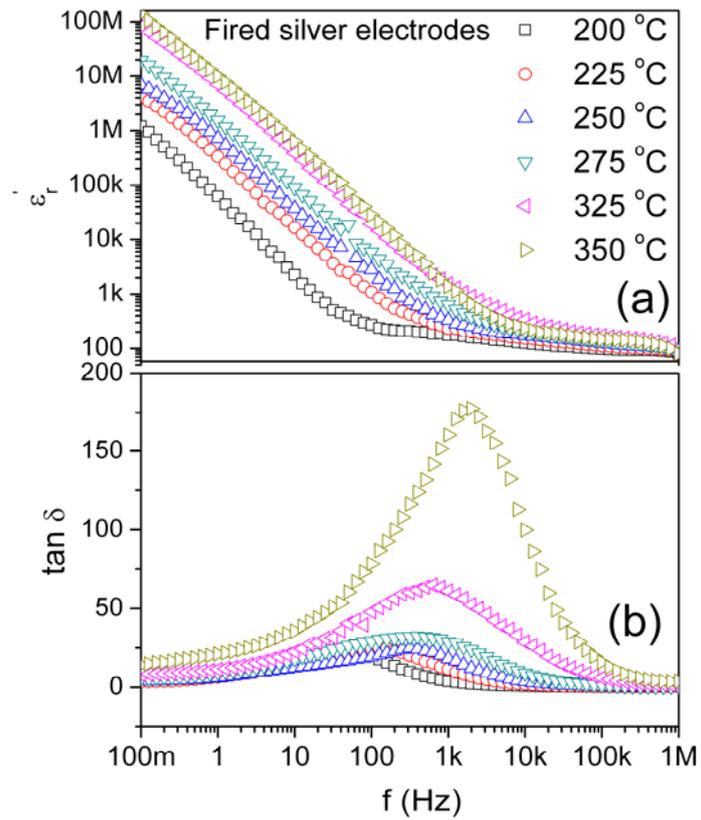

**Figure 8.**

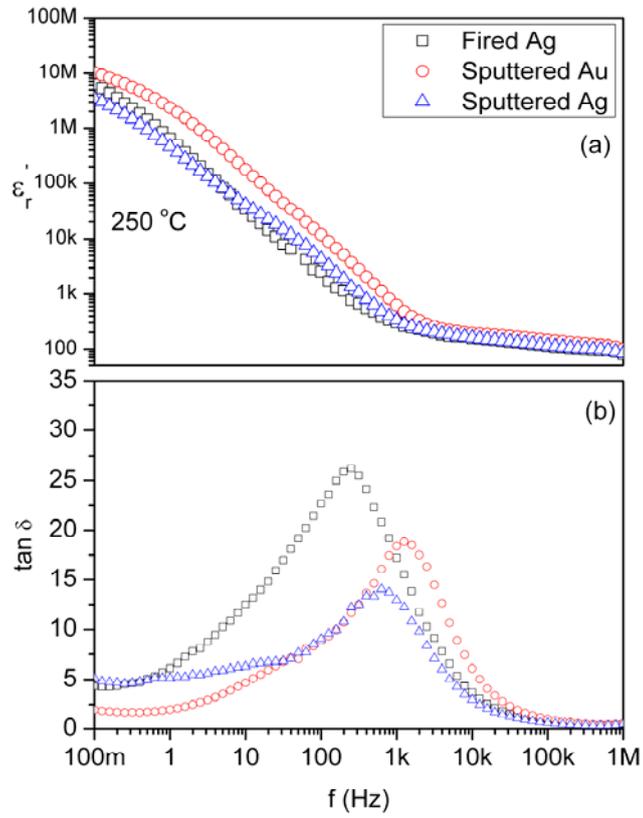

**Figure 9.**

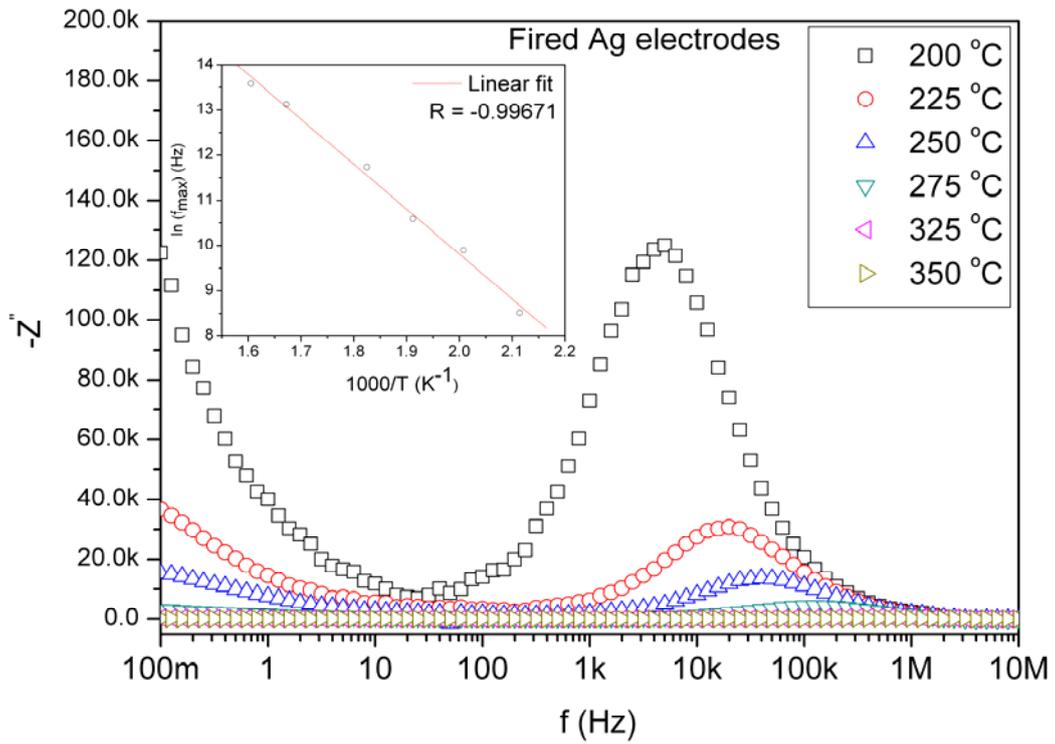

**Figure 10.**

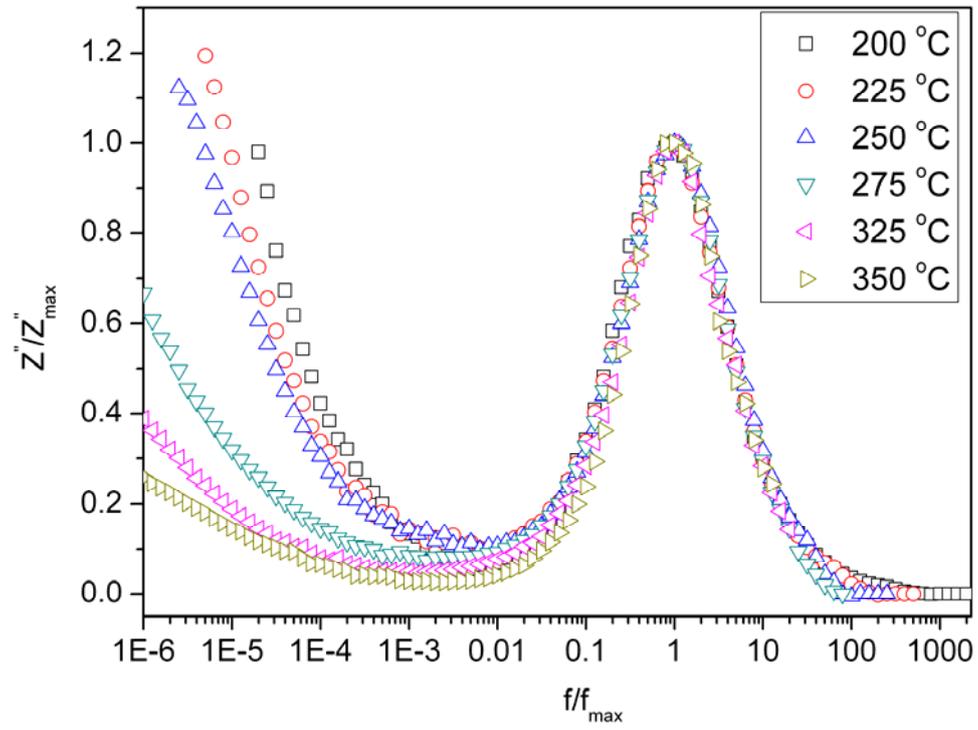

**Figure 11.**

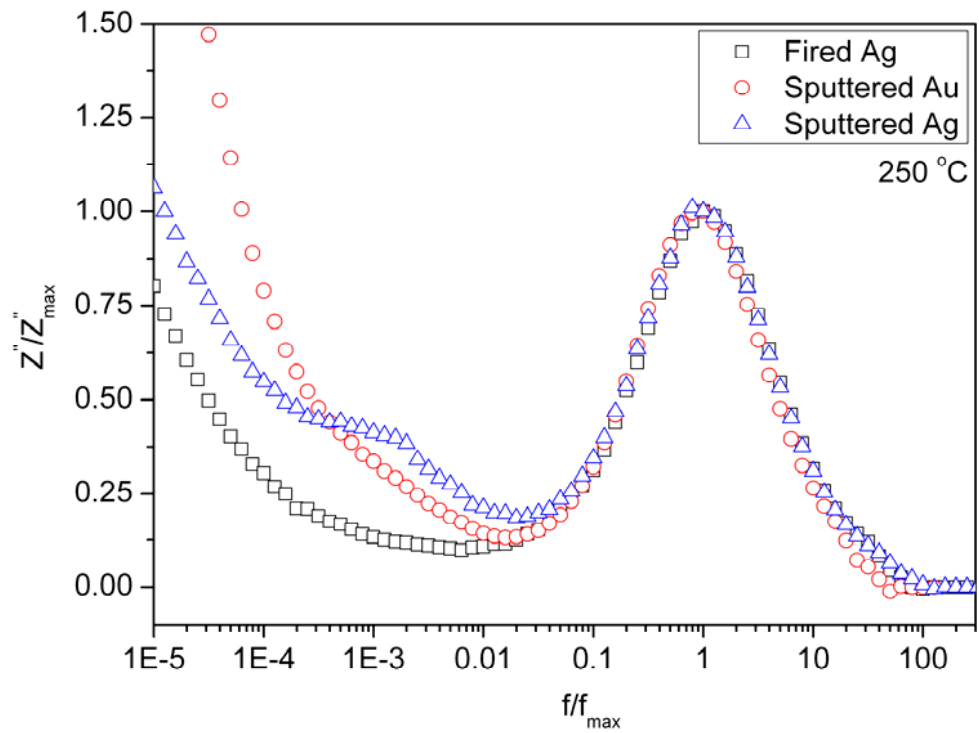

**Figure 12.**

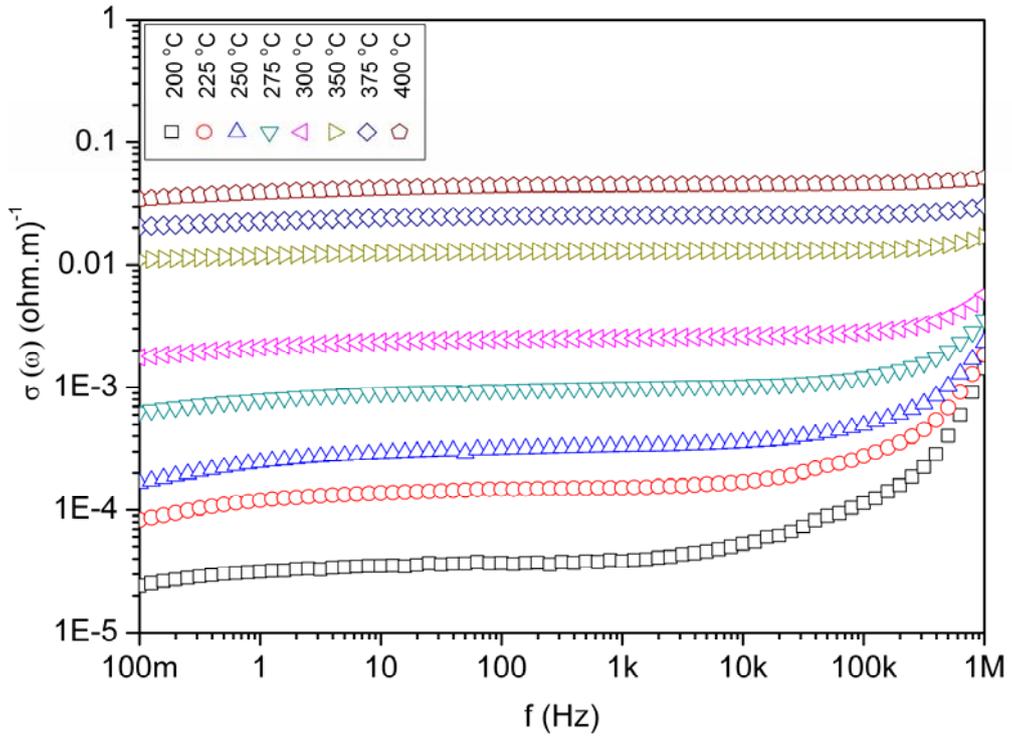

**Figure 13.**

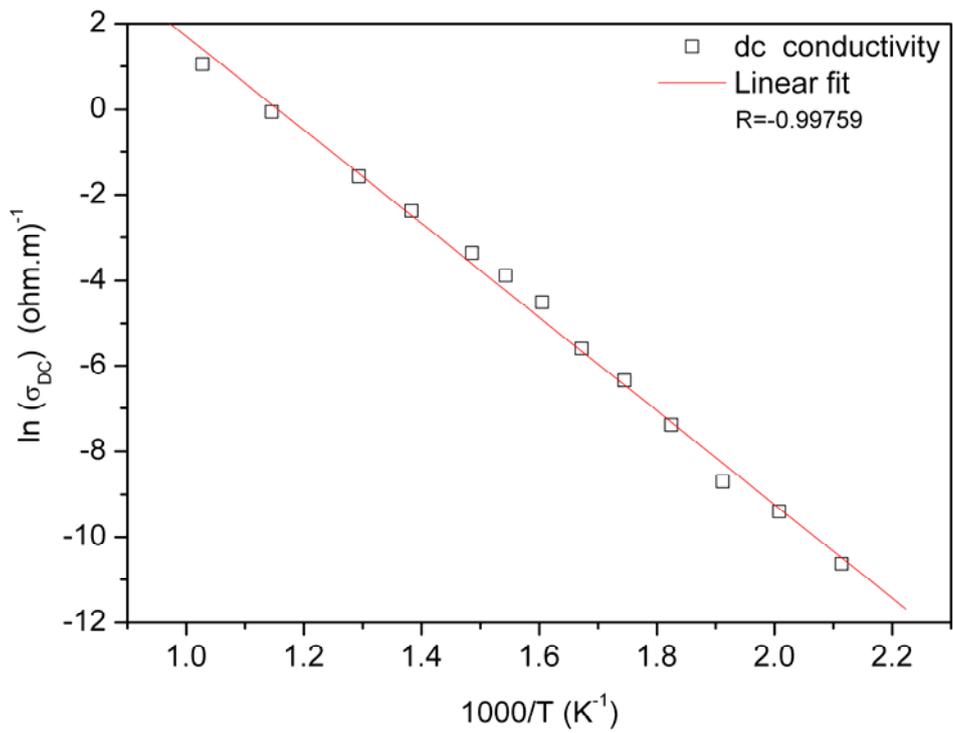